\documentclass
[prl,notitlepage,10pt,twocolumn,tightenlines,preprint,balancelastpage,floats,final]{revtex4}%
\usepackage{amsmath}
\usepackage{amsfonts}
\usepackage{amssymb}
\usepackage{graphicx}%
\providecommand{\U}[1]{\protect\rule{.1in}{.1in}}
\begin{document}
\preprint{ }
\title[Short title for running header]{Experimental joint weak measurement on a photon pair as a probe of Hardy's Paradox}
\author{J. S. Lundeen}
\author{A. M. Steinberg}
\affiliation{Department of Physics, University of Toronto, 60 St. George Street, Toronto
ON, M5S 1A7, Canada}

\begin{abstract}
It has been proposed that the ability to perform joint weak measurements on
post-selected systems would allow us to study quantum paradoxes. These
measurements can investigate the history of those particles that contribute to
the paradoxical outcome. Here, we experimentally perform weak measurements of
joint (i.e. nonlocal) observables. In an implementation of Hardy's Paradox, we
weakly measure the locations of two photons, the subject of the conflicting
statements behind the Paradox. Remarkably, the resulting weak probabilities
verify all these statements but, at the same time, resolve the Paradox.

\end{abstract}
\volumeyear{year}
\volumenumber{number}
\issuenumber{number}
\eid{identifier}
\date[Date text]{date}
\received[Received text]{date}

\revised[Revised text]{date}

\accepted[Accepted text]{date}

\published[Published text]{date}

\maketitle

Retrodiction is a controversial topic in quantum mechanics \cite{Hosten2006}.
How much is one allowed to say about the history (e.g. particle trajectories)
of a post-selected ensemble? Historically this has been deemed a question more
suitable for philosophy (e.g. counterfactual logic) than physics; since the
early days of quantum mechanics, the standard approach has been to restrict
the basis of our physical interpretations to direct experimental observations.
On the practical side of the question, post-selection has recently grown in
importance as a tool in fields such as quantum information: e.g. in linear
optics quantum computation (LOQC) \cite{Knill2001}, where it drives the logic
of quantum gates; and in continuous variable systems, for entanglement
distillation \cite{Browne2003}. Weak measurement is a relatively new
experimental technique for tackling just this question. It is of particular
interest to carry out weak measurements of multi-particle observables, such as
those used in quantum information. Here, we present an experiment that uses
weak measurement to examine the two-particle retrodiction paradox of Hardy
\cite{Hardy1992, Aharonov2002}, confirming the validity of certain
retrodictions and identifying the source of the apparent contradiction.

Hardy's Paradox is a contradiction between classical reasoning and the outcome
of standard measurements on an electron E and positron P in a pair of
Mach-Zehnder interferometers (see Fig. 1). Each interferometer is first
aligned so that the incoming particle always leaves through the same exit
port, termed the \textquotedblleft bright\textquotedblright\ port B (the other
is the \textquotedblleft dark\textquotedblright\ port D). The interferometers
are then arranged so that one arm (the "Inner" arm I) from each interferometer
overlaps at Y. It is assumed that if the electron and positron simultaneously
enter this arm they will collide and annihilate with 100\% probability. This
makes the interferometers \textquotedblleft Interaction-Free
Measurements\textquotedblright\ (IFM) \cite{Elitzur1993}: that is, a click at
the dark port indicates the interference was disturbed by an object located in
one of the interferometer arms, without the interfering particle itself having
traversed that arm. Therefore, in Hardy's Paradox a click at the dark port of
the electron (positron) indicates that the positron (electron) was in the
Inner arm. Consider if one were to detect both particles at the dark ports. As
IFMs, these results would indicate the particles were simultaneously in the
Inner arms and, therefore should have annihilated. But this is in
contradiction to the fact that they were actually detected at the dark ports.
Paradoxically, one does indeed observe simultaneous clicks at the dark ports
\cite{Irvine2005}, just as quantum mechanics predicts.

Weak measurements have been performed in classical optical experiments
\cite{Ritchie1991}, as well as on the polarization of single photons
\cite{Pryde2005}. Weak measurements of joint observables are particularly
important, as this class of observables includes nonlocal observables, which
can be used to create and identify multiparticle entanglement (e.g. in cluster
state computing \cite{Raussendorf2001}). Joint observables also include
sequential measurements on a single particle, allowing them to characterize
time-evolution in a system \cite{Aharonov2007}. In this experiment, we
demonstrate a new technique that for the first time enables us to perform
joint weak measurements. With this technique, we implement a proposal by
Aharonov et al. \cite{Aharonov2002} to weakly measure the simultaneous
location of the two path-entangled photons in Hardy's Paradox \cite{Hardy1992}%
. This technique opens up the possibility of in situ interrogation and
characterization of complex multiparticle quantum systems such as those used
in quantum information.%
%TCIMACRO{\FRAME{ftbpFU}{2.8608in}{2.7311in}{0pt}{\Qcb{Hardy's Paradox setup. E
%and P indicate electron and positron, respectively, which can collide in
%region H. BS1P, BS1E, BS2P, and BS2E are 50:50 beamsplitters. D and B are the
%dark and bright ports of the interferometers. I and O are the Inner and Outer
%interferometer arms.}}{}{Figure1.eps}{\special{ language "Scientific Word";
%type "GRAPHIC";  maintain-aspect-ratio TRUE;  display "USEDEF";
%valid_file "F";  width 2.8608in;  height 2.7311in;  depth 0pt;
%original-width 5.1526in;  original-height 4.9147in;  cropleft "0";
%filename 'Figure1.eps';file-properties "XNPEU";}}}%
%BeginExpansion
\begin{figure}
[ptb]
\begin{center}
\includegraphics[
height=2.7311in,
width=2.8608in
]%
{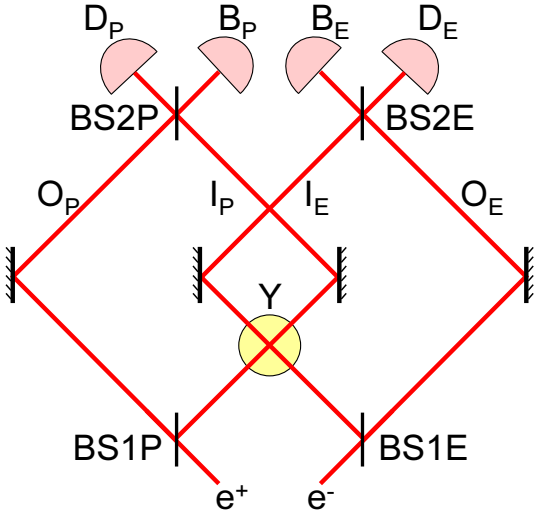}%
\caption{Hardy's Paradox setup. E and P indicate electron and positron,
respectively, which can collide in region H. BS1P, BS1E, BS2P, and BS2E are
50:50 beamsplitters. D and B are the dark and bright ports of the
interferometers. I and O are the Inner and Outer interferometer arms.}%
\end{center}
\end{figure}
%EndExpansion

A standard measurement collapses the measured system, irreversibly destroying
the original quantum state of the system. Post-selected subensembles are
particularly difficult to investigate since measurements on the ensemble
before the post-selection will collapse the system and, thus, alter the action
of the post-selection itself. Weak measurement was devised by Aharonov,
Albert, and Vaidman as a way of circumventing these problems
\cite{Aharonov1988}. It is an extension of the standard von Neumann
measurement model \cite{Von1955} in which the coupling $g$ between the
measured system and the measurement device is made asymptotically small. This
has the drawback of reducing the amount of information one retrieves in a
single measurement. The reward is that the consequent disturbance of the
measured system is correspondingly small. To extract useful information, one
must repeat the measurement on a large ensemble of identical quantum systems.
The average result is called the \textquotedblleft weak
value\textquotedblright, denoted $\left\langle \hat{C}\right\rangle _{W}$,
where $\hat{C}$ is the measured operator.

To set up Hardy's Paradox we use two photons instead of the electron and
positron. The experimental setup is shown in Fig. 2. A diode laser produces a
30 mW 405 nm beam (blue dashed line) which is filtered by a blue glass filter
(BF) and sent through a dichroic mirror (DM). This beam produces 810 nm photon
pairs (red solid line) in a 4 mm long BBO crystal through the process of Type
II spontaneous parametric downconversion. These pairs, consisting of a
horizontal (E)\ photon and a vertical (P) photon, take the place of the
electron and positron. The pump passes through a second DM, to later be
retroreflected. The photon pair passes through a filter (F) to remove any
residual pump light, followed by a 2mm thick BBO crystal (CC), to compensate
for the birefringent delay in the first crystal. The photon pair then meets a
50/50 beamsplitter (BS1EP), which acts as the first beamsplitter in both the E
and P interferometers, so that each photon can either be retroreflected and
enter the Inner arm or be transmitted and enter the Outer arm.%
%TCIMACRO{\FRAME{ftbpFU}{3.103in}{2.2053in}{0pt}{\Qcb{The experimental setup.
%Labels are explained in the text.}}{}{Figure2.eps}%
%width 3.103in;  height 2.2053in;  depth 0pt;  original-width 6.8346in;
%original-height 4.8369in;  cropleft "0";  croptop "1";  cropright "1";
%cropbottom "0";  filename 'Figure2.eps';file-properties "XNPEU";}} }%
%BeginExpansion
\begin{figure}
[ptb]
\begin{center}
\includegraphics[
height=2.2053in,
width=3.103in
]%
{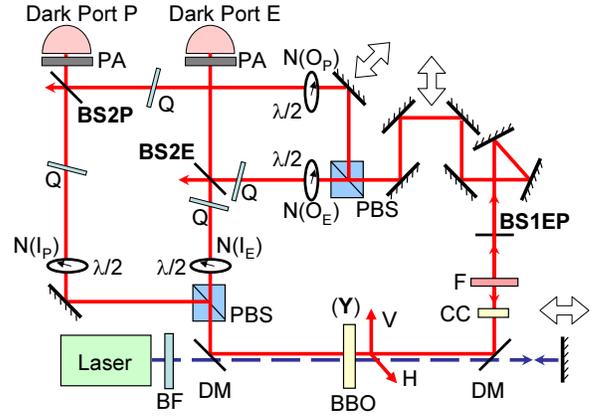}%
\caption{The experimental setup. Labels are explained in the text.}%
\end{center}
\end{figure}
%EndExpansion

In place of electron-positron annihilation, a quantum interference effect acts
as an absorptive two-photon switch (Y) \cite{Resch2001a}. Photons reflected
into the Inner arm pass back through the BBO crystal along with the
retroreflected pump beam. The amplitude for the retroreflected pump to create
a pair of photons in the crystal is set to interfere destructively with
amplitude for a photon pair in Inner arms.\ Thus, if both the E and P photons
enter their Inner arms they are removed, whereas if only a single photon
enters, it passes through the crystal unimpeded.

Photons transmitted at the first beamsplitter enter the Outer arms, which
contain a variable delay. Next, both the Inner and Outer paths encounter
polarizing beamsplitters (PBS) so that the E and P photons are split into
their own spatially separate interferometers. The P interferometer contains an
additional variable delay so that both interferometers can be adjusted to have
the same path-length difference. The Inner and Outer paths of the two
interferometers are recombined at two non-polarizing beamsplitters (BS2E and
BS2P), taking the place of the final Mach-Zehnder beamsplitters for the
electron and positron. Tilted quartz pieces (Q) before and after the NPBSs
compensate for undesired polarization phase-shifts in them.

Placing a half-waveplate in an arm allows us to measure whether a photon
travelled through this arm. To understand how this functions consider a
half-waveplate placed in the $E$ Outer arm aligned so as rotate the
polarization of a photon passing through it by 90$%
%TCIMACRO{\U{b0}}%
%BeginExpansion
{{}^\circ}%
%EndExpansion
$. The polarization of the photon arriving at the $E$ dark port then perfectly
indicates if it was in the $E$ Outer arm or not. This a measurement of the
\textquotedblleft occupation\textquotedblright\ $\hat{N}\left(  M_{K}\right)
$ of the $M=I$ or $O$ (Inner or Outer) interferometer arm by photon $K=E$ or
$P$ (e.g. $\hat{N}\left(  O_{E}\right)  =\left\vert O_{E}\right\rangle
\left\langle O_{E}\right\vert $). Unfortunately this procedure is a standard
projective measurement and, hence strongly disturbs the system. In particular,
the interference will be destroyed as the two paths are now completely
distinguishable and, thus, the interferometer will not function as an IFM. The
strength of the measurement interaction ($\hat{U}=\exp\left(  -ig\hat{N}%
\hat{\sigma}_{y}\right)  $) is parameterized by $g\approx\theta,$ the
polarization rotation. In this experiment, we reduce this disturbance by
rotating the photon's polarization by only 20$%
%TCIMACRO{\U{b0}}%
%BeginExpansion
{{}^\circ}%
%EndExpansion
$, reducing $g$ four-fold, and thereby performing a weak measurement. The
trade-off is that it is now impossible to know which arm a particular detected
photon went through. Instead, we measure the average polarization rotation at
the detector over many trials to find what fraction of photons passed through
that particular arm. If no rotation is observed then the classical inference
would be that the photon was never in the arm with the waveplate. Conversely,
if we measure an average rotation of 20$%
%TCIMACRO{\U{b0}}%
%BeginExpansion
{{}^\circ}%
%EndExpansion
$ one might infer that every photon passed through the waveplate. Quantum
mechanically, this rotation constitutes a weak measurement of the occupation
$\hat{N}$ of a particular interferometer arm.

The crux of the paradox is that the detected photons cannot have
simultaneously been in the Inner arms. To test this we require a
weak-measurement of the \textit{joint} occupation of two arms. It was
previously thought that a physical interaction between the particles was
necessary to make weak-measurements of joint observables (e.g. the
electrostatic interaction of ions, as in Ref. \cite{Molmer2001}). In Refs.
\cite{Resch2004a}, we theoretically showed that one only needs to perform
single-particle weak measurements on each particle. The joint weak values then
appear in polarization correlations between the two particles as follows:
\begin{align}
\left\langle \hat{N}(M_{K})\right\rangle _{W}  &  =g^{-1}\operatorname{Re}%
\left\langle \hat{\sigma}_{zK}^{-}\right\rangle \\
\left\langle \hat{N}(M_{E})\hat{N}(M_{P})\right\rangle _{W}  &  =g^{-2}%
\operatorname{Re}\left\langle \hat{\sigma}_{zE}^{-}\hat{\sigma}_{zP}%
^{-}\right\rangle , \label{jointweak}%
\end{align}
where $\hat{\sigma}_{zK}^{-}=(\hat{\sigma}_{xK}-i\hat{\sigma}_{yK})$ is the
z-basis lowering operator for the polarization of photon $K=E$ or $P$. In
practice, we independently measure $g$ for each arm to account for
polarization-dependent losses. We weakly measure all four combinations of
$\hat{N}(M_{E})\hat{N}(M_{P})$ by placing half-waveplates ($\lambda$/2) in all
four arms just before the final beamsplitters. We measure the occupation of a
particular pair of arms by rotating only those two waveplates. After the final
beamsplitters we measure average polarization rotations as well as the
correlations specified in Eq. \ref{jointweak} with polarization analyzers (PA)
consisting of a quarter-waveplate and polarizer followed by a single-photon
detector (Perkin Elmer SPCM-AQR). Once the Pauli operators are substituted in
Eq. \ref{jointweak} and the real part is found, four Pauli operators remain in
the final expectation value. For each of these Pauli operators, the analyzer
must be set to two positions (e.g. for $\hat{\sigma}_{x}$, $45^{\circ}$ and
$-45^{\circ}$ ($\nearrow$ and $\nwarrow$) and for $\hat{\sigma}_{y}$,
right-hand circular and left-hand circular ($\circlearrowright$ and
$\circlearrowleft$)). Thus, each joint weak value requires eight measurements
of coincidence rates at the two dark ports:
\begin{multline}
\operatorname{Re}\left\langle \hat{\sigma}_{zE}^{-}\hat{\sigma}_{zP}%
^{-}\right\rangle =\frac{R_{\nearrow\nearrow}+R_{\nwarrow\nwarrow}%
-R_{\nwarrow\nearrow}-R_{\nearrow\nwarrow}}{R_{\nearrow\nearrow}%
+R_{\nwarrow\nwarrow}+R_{\nwarrow\nearrow}+R_{\nearrow\nwarrow}}\\
-\frac{R_{\circlearrowright\circlearrowright}+R_{\circlearrowleft
\circlearrowleft}-R_{\circlearrowleft\circlearrowright}-R_{\circlearrowright
\circlearrowleft}}{R_{\circlearrowright\circlearrowright}+R_{\circlearrowleft
\circlearrowleft}+R_{\circlearrowleft\circlearrowright}+R_{\circlearrowright
\circlearrowleft}},
\end{multline}
where $R_{sq}$ is the coincidence rate when the $P(E)$ analyzer is set to
$s(q)$. Single weak values for the occupation of photon $E$ $(P)$ are found
from these rates by summing over analyzer settings for photon $P$ $(E)$. As an
example, we give the measurements contributing to $\left\langle \hat{N}%
(O_{E})\hat{N}(O_{P})\right\rangle _{W}$ : $R_{\nearrow\nearrow}%
=556,R_{\nwarrow\nwarrow}=583,R_{\nwarrow\nearrow}=834,R_{\nearrow\nwarrow
}=730,R_{\circlearrowright\circlearrowright}=571,R_{\circlearrowleft
\circlearrowleft}=543,R_{\circlearrowleft\circlearrowright}%
=666,R_{\circlearrowright\circlearrowleft}=750$ (all in counts per 420s) and
$g^{2}=0.365$.

In Table 1 we present the weak values for the various arm occupations. The
bottom cells and rightmost cells give the weak value for the occupation of a
single arm and the inner cells give the joint occupation of a pair of arms.
Error bars are derived from uncertainties in $g$ and statistical variations in
the rates\medskip.\newline$\medskip%
\begin{tabular}
[c]{|c|c|c||c|}\hline
& $N(I_{P})$ & $N(O_{P})$ & \\\hline
$N(I_{E})$ & $%
\begin{array}
[c]{c}%
0.245\pm0.068\\
\left[  0\right]
\end{array}
$ & $%
\begin{array}
[c]{c}%
0.641\pm0.083\\
\left[  1\right]
\end{array}
$ & $%
\begin{array}
[c]{c}%
0.926\pm0.015\\
\left[  1\right]
\end{array}
$\\\hline
$N(O_{E})$ & $%
\begin{array}
[c]{c}%
0.719\pm0.074\\
\left[  1\right]
\end{array}
$ & $%
\begin{array}
[c]{c}%
-0.759\pm0.083\\
\left[  -1\right]
\end{array}
$ & $%
\begin{array}
[c]{c}%
-0.078\pm0.02\\
\left[  0\right]
\end{array}
$\\\hline\hline
& $%
\begin{array}
[c]{c}%
0.924\pm0.024\\
\left[  1\right]
\end{array}
$ & $%
\begin{array}
[c]{c}%
0.087\pm0.023\\
\left[  0\right]
\end{array}
$ & \\\hline
\end{tabular}
$\newline Table 1. The weak values for the arm occupations in Hardy's
Paradox.\medskip

Examining the table reveals that the single-particle weak measurements are
consistent with the clicks at each dark port; as the IFM results imply, the
weakly measured occupations of each of the Inner arms are close to one and
those of each of the Outer arms are close to zero. The weak measurements
indicate that, at least when considered individually, the photons were in the
Inner arms. However, if we instead examine the joint occupation of the two
Inner arms, it appears that the two photons are only simultaneously present
roughly one quarter of the time. This demonstrates that, as we expect, the
particles are not in the inner arms together.

So far, we seem to have confirmed both of the premises of Hardy's Paradox: to
wit, that when $D_{P}$ and $D_{E}$ fire, $N(I_{P})$ and $N(I_{E})$ are close
to one (since the IFMs indicate the presence of the particles in Y) -- but
that $N(I_{P}\&I_{E})$ is close to zero (since when \textit{both} particles
are in Y, they annihilate and should not be detected). This is odd because in
classical logic, $N(I_{P}\&I_{E})$ must be $\geq N(I_{P})+N(I_{E})-1$; this
inequality is violated by our results. Although $N(I_{E})$ is $93\%$ and
$N(I_{P})$ is $92\%$, the data in Table 1 suggest that when E is in the Inner
path, P is not, and vice versa; hence the large values for $N(I_{E}%
\&O_{P})=64\%$ and $N(O_{E}\&I_{P})=72\%$. The fact that the sum of these two
seemingly disjoint joint-occupation probabilities exceeds 1 is the
contradiction with classical logic. In the context of weak measurements, the
resolution of this problem lies in the fact that weak valued probabilities are
not required to be positive definite \cite{Aharonov2002}, and so a negative
occupation $N(O_{E}\&O_{P})=-76\%$ is possible, preserving the probability sum
rules. In an ideal implementation of Hardy's Paradox, the joint probabilities
are strictly $0$ for both particles to be in Inner arms, $-1$ for both to be
in the Outer, and 1 for either to be in the Inner while the other is in the
Outer arm. These are indicated in brackets in Table 1, for comparison with our
experimental data. Discrepancies are because of the imperfect switch
efficiency ($85\pm3\%$) and IFM\ probabilities ($95\pm3\%$ for the E IFM and
$94\pm4\%$ for P).

What is the meaning of the negative joint occupation? Recall that the joint
values are extracted by studying the polarization rotation of both photons in
coincidence. Consider a situation in which both photons always simultaneously
passed through two particular arms. When a polarization rotator is placed in
each of these arms it would tend to cause their polarizations to rotate in a
correlated fashion; when P was found to have $45%
%TCIMACRO{\unit{\U{b0}}}%
%BeginExpansion
\operatorname{{{}^\circ}}%
%EndExpansion
$ polarization, E would also be more likely to be found at $45%
%TCIMACRO{\unit{\U{b0}}}%
%BeginExpansion
\operatorname{{{}^\circ}}%
%EndExpansion
$ than $-45%
%TCIMACRO{\unit{\U{b0}}}%
%BeginExpansion
\operatorname{{{}^\circ}}%
%EndExpansion
$. Experimentally, we find the reverse -- when P is found to have $45%
%TCIMACRO{\unit{\U{b0}}}%
%BeginExpansion
\operatorname{{{}^\circ}}%
%EndExpansion
$ polarization, E is preferentially found at $-45%
%TCIMACRO{\unit{\U{b0}}}%
%BeginExpansion
\operatorname{{{}^\circ}}%
%EndExpansion
$ (and vice versa), as though it had rotated in the direction opposite to the
one induced by the physical waveplate. As in all weak measurement experiments,
a negative weak value implies that the shift of a physical \textquotedblleft
pointer\textquotedblright\ (in this case, photon polarization) has the
opposite sign from the one expected from the measurement interaction itself.

In summary, Hardy's Paradox is a set of conflicting classical logic statements
about the location of the particles in each of two Mach-Zehnder
interferometers. It is impossible to simultaneously verify these statements
with standard measurements since testing one statement disturbs the system and
consequently nullifies the other statements. We attempt to minimize this
disturbance by reducing the strength of the interaction used to perform the
measurement. The results of these weak measurements indicate that all the
logical statements are correct and also provide a self-consistent, if strange,
resolution to the paradox. Since they do not disturb subsequent post-selection
of the systems under study, weak measurements are ideal for the interrogation
and characterization of post-selected multiparticle states such as GHZ or
Cluster states, and processes such as Linear Optics Quantum Computation. This
experiment demonstrates a new technique that, for the first time, allows for
the weak measurement of general multiparticle observables in these systems.

This work was supported by NSERC, QuantumWorks, CIPI, and the Canadian
Institute for Advanced Research. We thank Kevin Resch and Morgan Mitchell for
helpful discussions.

\end{document}